\documentclass[graybox]{svmult}

\usepackage{mathptmx}       
\usepackage{helvet}         
\usepackage{courier}        
\usepackage{type1cm}        
\usepackage{makeidx}         
\usepackage{graphicx}        
\usepackage{multicol}        
\usepackage[bottom]{footmisc}

\makeindex             

\begin{document}

\title*{Magnetohydrodynamic jets from different magnetic field configurations}
\titlerunning{MHD jets from different field distributions}
\author{Christian Fendt}
\authorrunning{Ch.~Fendt}
\institute{Christian Fendt \at 
           Max Planck Institute for Astronomy, K\"onigstuhl 17, D-69117 Heidelberg, 
           \email{fendt@mpia.de}}
%
%
\maketitle

\abstract{
Using axisymmetric MHD simulations we investigate how the overall jet formation is
affected by a variation in the disk magnetic flux profile and/or the existence of 
a central stellar magnetosphere.
Our simulations evolve from an initial, hydrostatic equilibrium state in a
force-free magnetic field configuration.
We find a unique relation between the collimation degree and the disk wind
magnetization power law exponent.
The collimation degree decreases for steeper disk magnetic field profiles.
Highly collimated outflows resulting from a flat profile tend to be unsteady.
We further consider a magnetic field superposed of a stellar dipole and 
a disk field in parallel or anti-parallel alignment.
Both stellar and disk wind may evolve in a pair of outflows, however, 
a reasonably strong disk wind component is essential for jet collimation.
Strong flares may lead to a sudden change in mass flux by 
a factor two.
We hypothesize that such flares may eventually trigger jet knots.
}

\section{Jets as collimated MHD flows}
\label{sec:1}
Astrophysical jets are launched by magnetohydrodynamic (MHD)
processes in the close vicinity of the central object --
an accretion disk surrounding a protostar or a compact object
[1,2, 9, 20, 21, 24].
Numerical simulations of MHD jet formation are essential for 
our understanding of the physical processes involved.
In general, simulations may be distinguished in those
taking into account the evolution of the disk structure
and others considering the disk surface as a fixed-in-time
boundary condition for the jet.
The first approach allows to directly investigate the mechanism lifting 
matter from the disk into the outflow 
[3, 10, 11, 14, 17, 18, 22, 24] 
This approach is computationally expensive and still
somewhat limited by spatial and time resolution.
In order to study the acceleration and collimation of a disk/stellar
wind it is essential to follow
the dynamical evolution for i) very long time 
ii) on a sufficiently large grid with iii) appropriate resolution.
For such a goal, the second approach is better suited
[4-7, 12, 13, 15, 19, 25], allowing as well for parameter studies.
The case of superposed stellar/disk magnetic field is rarely
treated in simulations, still, the first model was discussed 
already in [23].
Simulations of a dipole with aligned vertical 
disk field are presented by  [16, 18].
The stellar field has important impact on the jet formation 
process as enhancing the magnetic flux, adding a central pressure,
and providing excess angular momentum for the launching region.

\begin{figure}[t]
\sidecaption
\includegraphics[scale=.19]{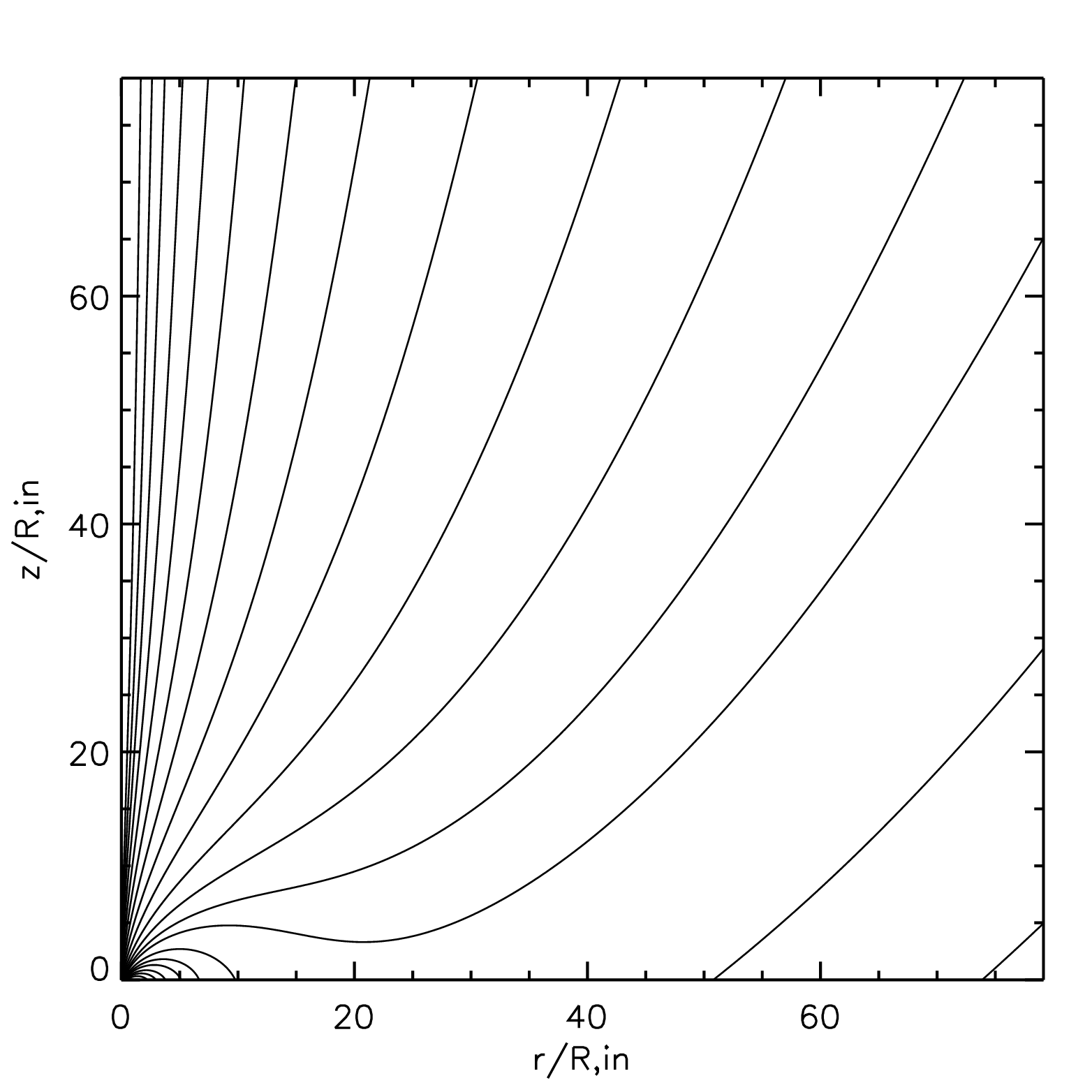}
\includegraphics[scale=.19]{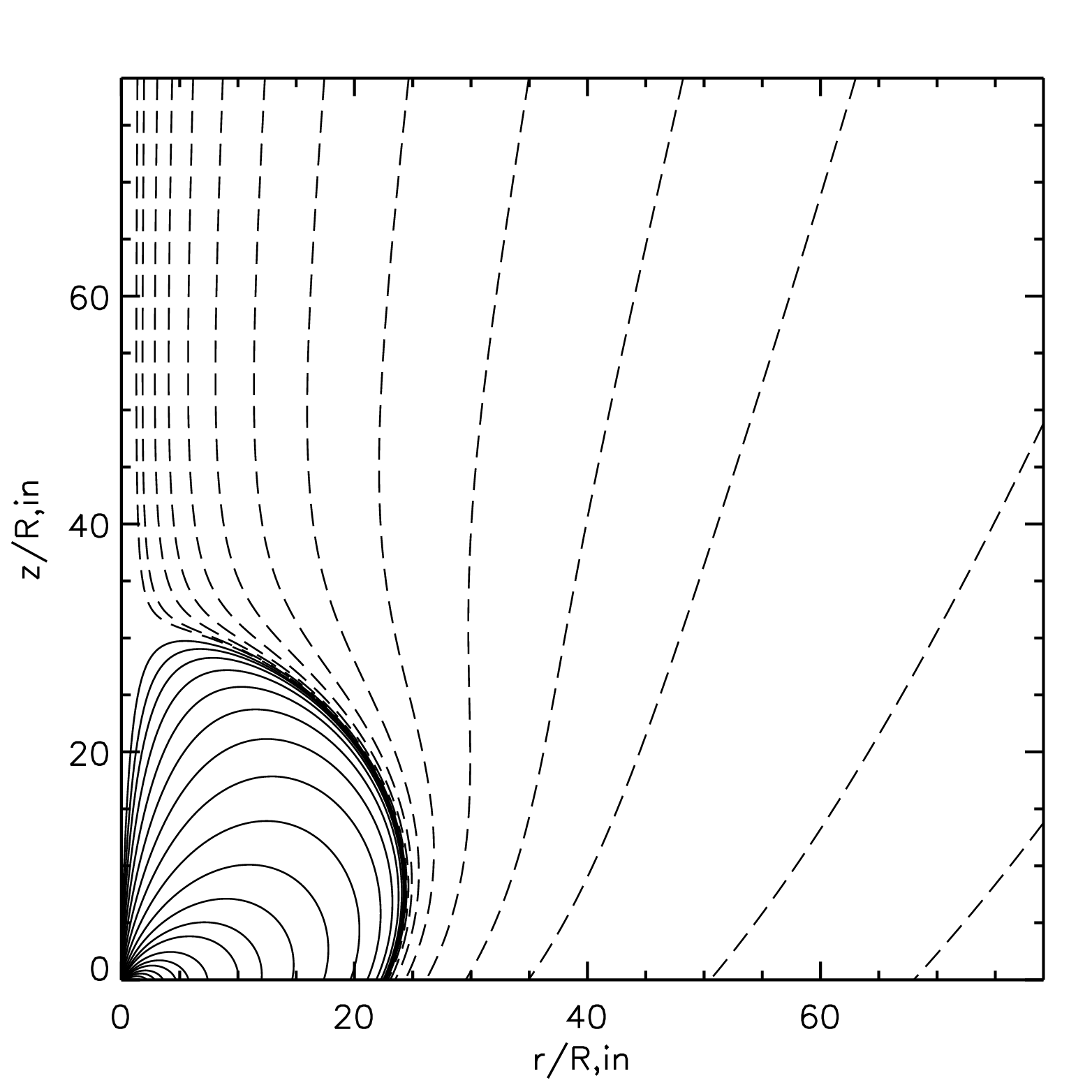}
\includegraphics[scale=.19]{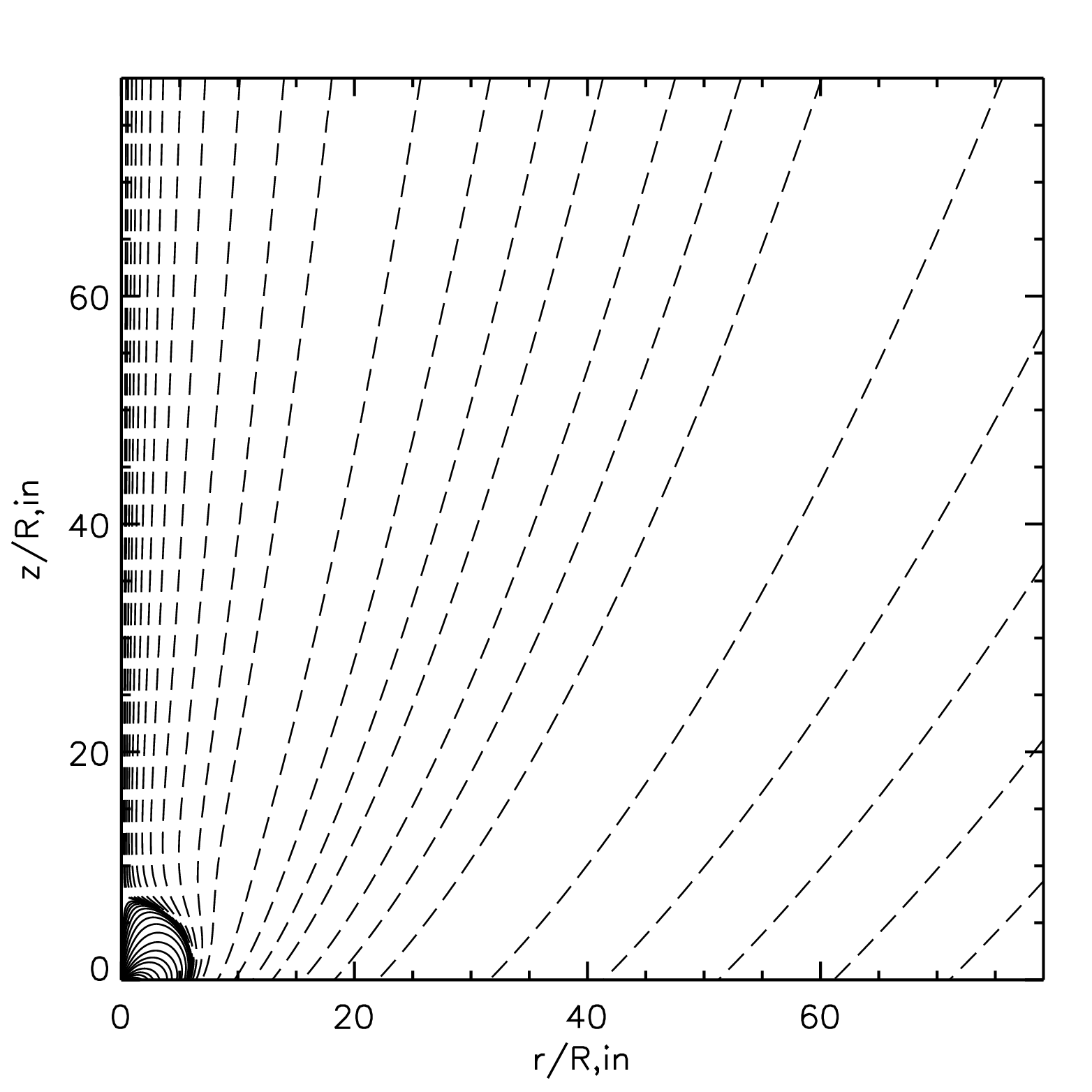}
%
%
\caption{
Example initial magnetic field distributions 
(poloidal magnetic field lines). 
Full and dashed lines indicate the direction of magnetic flux.
Magnetic field parameters:
    $\Psi_{0,\rm d} = 0.01, -0.01, -0.1, $ resp.
    $\Psi_{0, \star} = 5.0, 5.0, 3.0$ (from left to right). From [5].
}
\label{SDW-initial}       
\end{figure}

\section{Model setup}
We use the ZEUS-3D MHD code extended for physical magnetic resistivity
(see description in [6]).
The set of MHD equations considered is the following,
\begin{equation}
{\frac{\partial \rho}{\partial t}} + \nabla \cdot (\rho \vec{v} ) = 0,\,\,
\nabla \cdot\vec{B} = 0,\,\,
\frac{4\pi}{c} \vec{j} = \nabla \times \vec{B},\,\,
{\frac{\partial\vec{B} }{\partial t}}
- \nabla \times \left(\vec{v} \times \vec{B}
-{\frac{4\pi}{c}} \eta\vec{j}\right)= 0
\end{equation}
\begin{equation}
\rho \left[ {\frac{\partial\vec{u} }{\partial t}}
+ \left(\vec{v} \cdot \nabla\right)\vec{v} \right]
+ \nabla (p+p_A) +
 \rho\nabla\Phi - \frac{\vec{j} \times \vec{B}}{c} = 0,
\end{equation}
with the usual notation [4-7, 19].
We do not solve the energy equation, but apply an internal energy 
$ e=p/(\gamma-1)$ of a polytropic gas ($\gamma =5/3$).
Turbulent Alfv\'enic pressure $p_{\rm A}$ allows for a "cool" corona.
The turbulent magnetic diffusivity $\eta(r,z;t)$ can be related 
to $p_{\rm A}$ applying our toy model [6]. 
The $\eta \simeq 0.01$ was chosen low and does not affect collimation.
Diffusivity is, however, essential for reconnection processes.
We distinguish setup {\em DW} (pure disk wind) and {\em SDW}
(stellar wind plus disk wind) by choice of boundary and 
initial conditions.
Model DW investigates different disk magnetic field
and mass flux profiles [4].
Model SDW investigates the interrelation of the stellar magnetosphere 
with the surrounding disk jet [5].

\begin{figure}[t]
\sidecaption
\includegraphics[scale=.25]{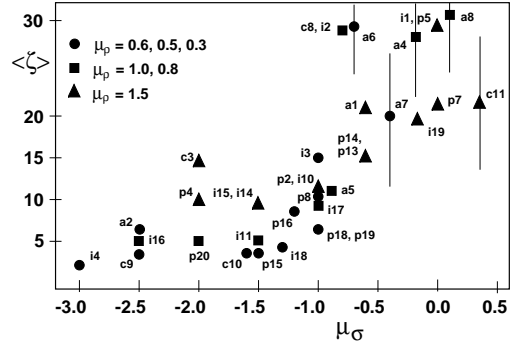}
%
%
\caption{
Collimation degree $<\zeta >$ and power index of the disk 
wind magnetization profile  $\mu_{\sigma }$.
Bars indicate simulations with time variable collimation degree (from [4]).
}
\label{sigma-coll}       
\end{figure}

{\em Boundary conditions:}
In setup DW we distinguish along the equatorial plane the gap region 
$r< 1.0$ and disk region $r>1.0$.
The magnetic field is fixed in time and is determined by the initial 
condition.
We have chosen a power law, $B_{\rm p}(r,0)\!\sim\! r^{-\mu}$, 
and investigate different $\mu$.
In setup SDW we further distinguish the star from $r=0 - 0.5$,
and the gap from $r=0.5 - 1.0$.
Co-rotation radius and inner disk radius coincide.
A Keplerian disk is the boundary condition for the mass inflow from 
the disk surface into the corona. 
Matter is ``injected'' from the disk (and the star) with low velocity
$\vec{v}_{\rm inj}(r,0) = \nu_{\rm i} v_{\rm K}(r) \vec{B}_{\rm P}/B_{\rm P}$
and density $\rho_{\rm inj}(r,0) = \eta_{\rm i}\,\rho(r,0)$.
Typically, $\nu_{\rm i} \simeq 10^{-3}$ and $\eta_{\rm i} \simeq 100$ for 
stellar and disk wind, but could be chosen differently.

{\em Initial conditions:}
As initial state we prescribe a force-free magnetic field and a
hydrostatic equilibrium $\rho(r,z, t=0) = (r^2 + z^2)^{-3/4}$.
For model DW we calculate the initial field distribution 
from the disk magnetic field profile using our finite
element code (see [4,8]).
For model SDW the initial field is a superposed dipole plus disk field.
For the disk component we apply the potential field of [6,19].
We prescribe the initial field by the magnetic flux distribution
$\Psi(r,z) \equiv \int \vec{B}_{\rm p} d\vec{A}$,
\begin{eqnarray}
\Psi(r,z) =
\Psi_{0,\rm d}\,\frac{1}{r}\left(\sqrt{r^2+(z_{\rm d}+z)^2}-(z_{\rm d}+z)\right)
 + \Psi_{0,\star}\,\frac{r^2}{\left(r^2+(z_{\rm d}+z)^2\right)^{3/2}}.
\end{eqnarray}
Certain field combinations are investigated,
parameterized by the disk $\Psi_{0,\rm d}$ and stellar magnetic flux $\Psi_{0,\star}$
(Fig.~\ref{SDW-initial}).

\section{Disk wind magnetization and jet collimation}
Simulations of setup DW were run for different disk magnetic field 
profiles $B_{\rm p}(r,z=0) \sim r^{-\mu}$ and 
density profiles $\rho_{\rm inj}(r,z=0) \sim r^{-\mu_{\rho}}$.
In general, we find an
increasing degree of collimation with decreasing slope of the disk magnetic field
profile (see [4]).
This seems to rule out launching models for collimated jets from a concentrated 
magnetic flux such as e.g. the X-wind scenario.
A steep density profile leads to a higher collimation degree, which is not 
surprising as the mass flux is more concentrated just by definition of the 
boundary condition.
A physically meaningful classification taking into account both density and 
magnetic field can be achieved by comparing the
degree of collimation degree versus disk wind {\em magnetization} profile,
$\sigma(r,z=0) \sim B_{\rm p}^2(r) r^4 \rho_{\rm inj}^{-1} v_{\rm inj}^{-1}(r)\Omega_K(r)^2$,
thus, 
$\sigma(r,z=0) \sim r^{-(2\mu - \mu_{\rho} +1/2)} \equiv r^{\mu_{\sigma}}$.
The resulting diagram Fig.~\ref{sigma-coll} shows a convincing correlation
between the magnetization power law index $\mu_{\sigma}$ and the average 
degree of collimation $<\!\!\zeta\!\!>$.
The width of the $(\mu_{\sigma}$ - $<\!\!\zeta\!\!>)$-correlation 
is due to further differences in the parameter space.

\begin{figure}[t]
\sidecaption
\includegraphics[scale=.18]{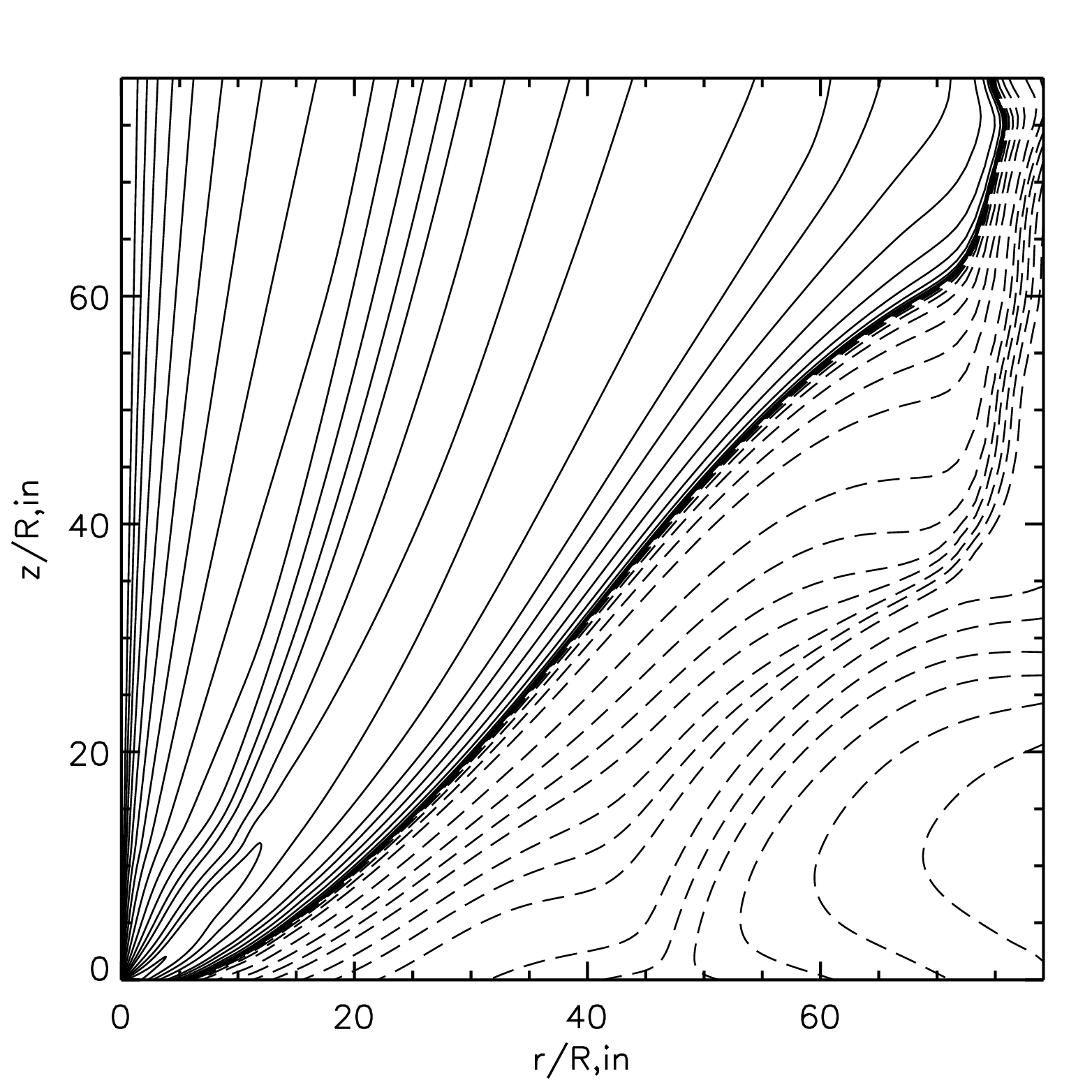}
\includegraphics[scale=.18]{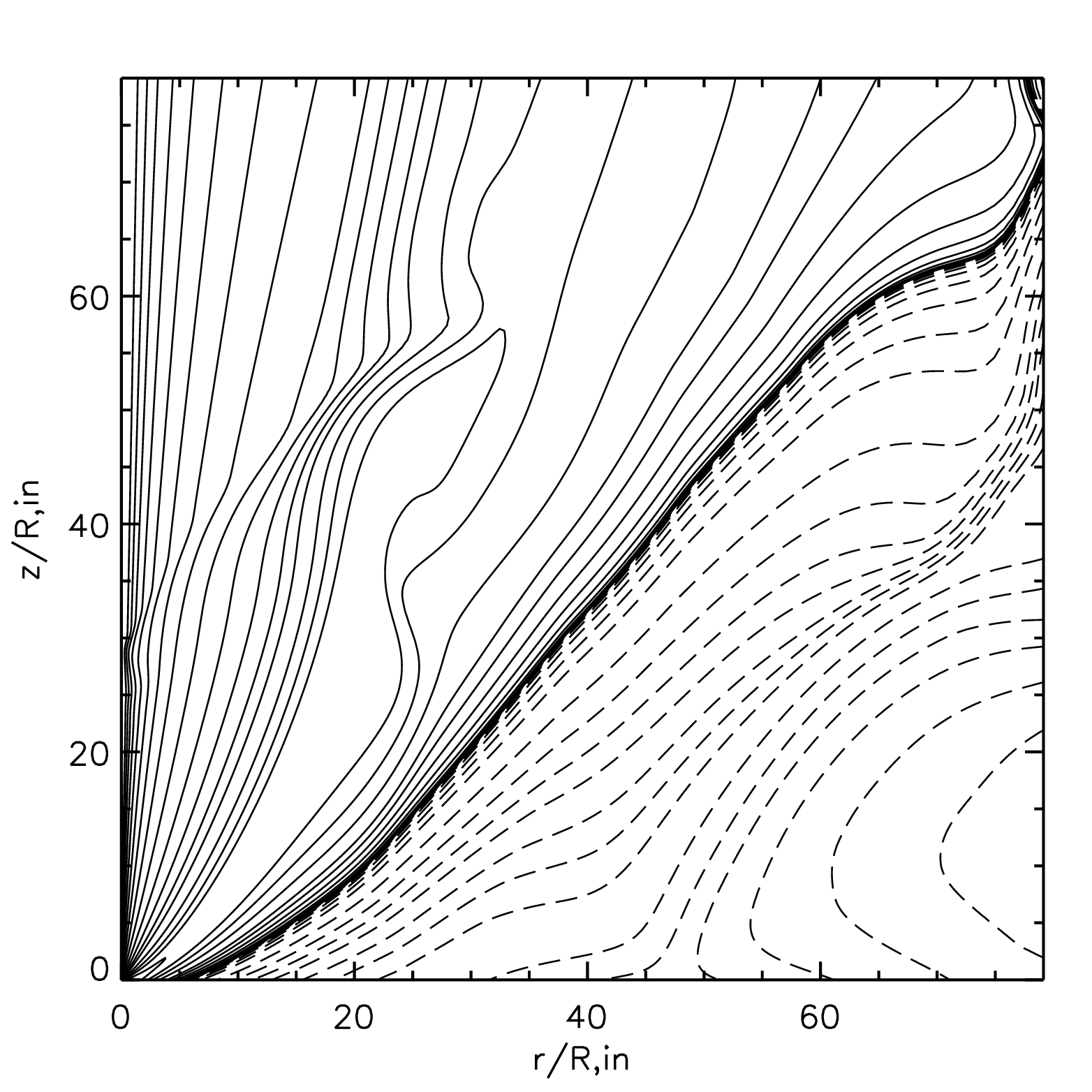}
\includegraphics[scale=.18]{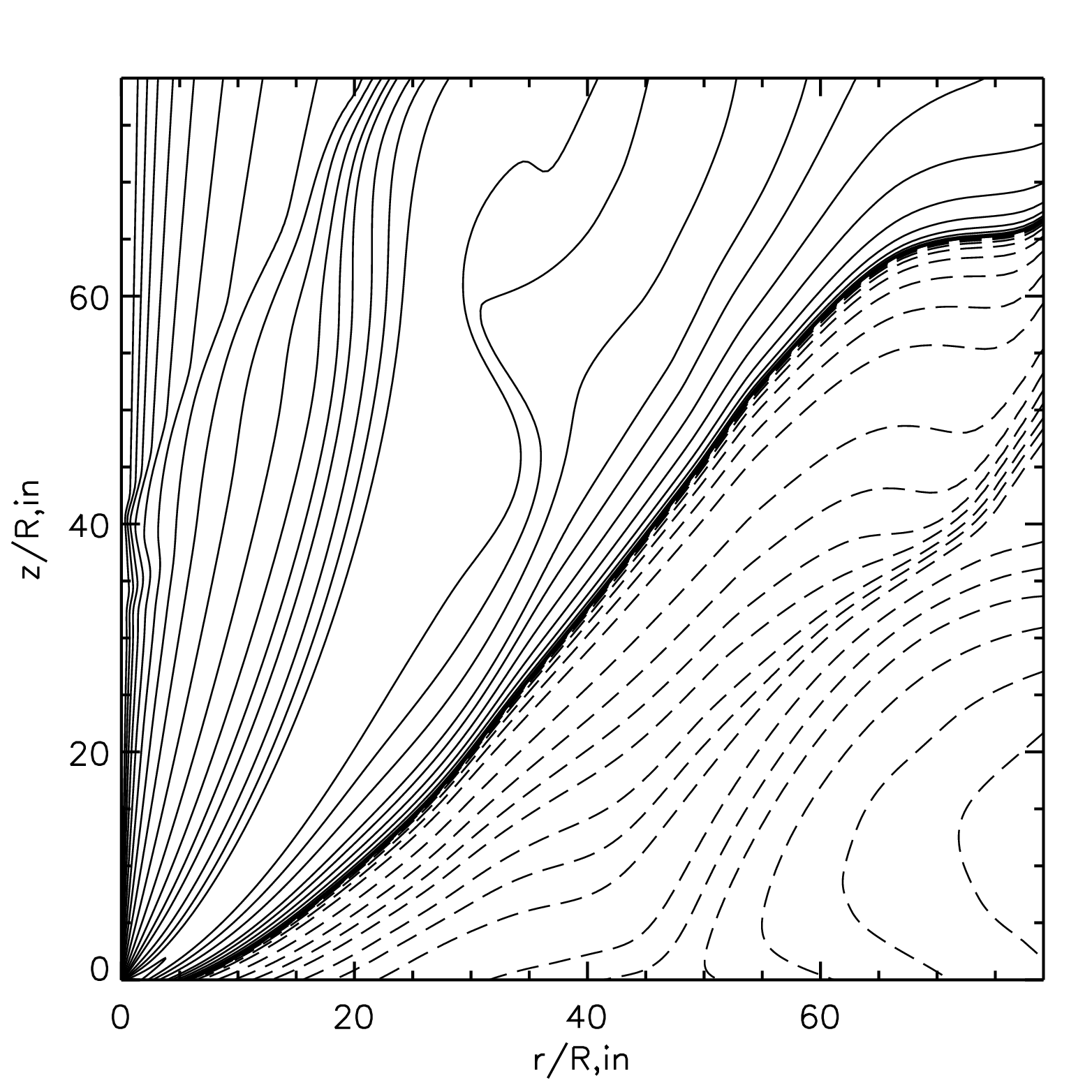}
%
%
\caption{
Poloidal magnetic field evolution during one example flare around $t=180$.
Solid and dashed lines indicate the direction of total magnetic flux of the
superposed dipolar and disk magnetic field components.
Shown are time steps: 1760, 1790, 1810 (from left to right).
}
\label{flare}       
\end{figure}

\section{Jet mass flux triggered by star-disk magnetospheric flares}
Simulations of setup SDW were run for aligned and anti-aligned orientation
of dipole versus disk field and for different strength of
both field contributions [5].

Independent of the alignment, the central dipole does not survive
on the large scale.
A two-component outflow emerges as stellar wind plus disk wind.
For a reasonably strong disk magnetic flux a collimated jet emerges.
If the overall outflow is dominated by a strong stellar outflow
a low mass flux disk wind remains uncollimated.
The best setup to launch a collimated jet from a star-disk magnetosphere 
is that of a relatively heavy disk wind and high disk magnetic flux.
Stellar wind dominated simulations may give a high degree of collimation,
however they collimate to too small radii.
Stellar magnetic flux dominated outflows tend to stay un-collimated.

\begin{figure}[t]
\sidecaption
\includegraphics[scale=.40]{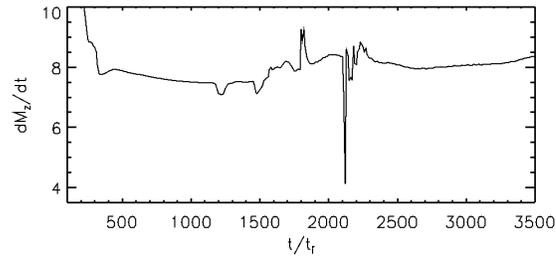}
%
%
\caption{Axial mass flux integrated along the upper $z$-boundary
versus time. Note the change of mass flux of $10-50\% $ during
the flare events. High mass fluxes for $ t<500 $ indicate sweeping off of
the initial corona.
}
\label{mflux}       
\end{figure}

In some simulations we observe reconnection flares, similar to
coronal mass ejections, typically expanding and reconnecting within 70 
orbital periods of the inner disk.
This is similar to [10], however, in their case reconnection is are 
triggered by time-variation of the accretion rate.
In our case the reconnection/flares seem to be triggered by the evolution 
of the outer disk wind.
Even for our very long time-scales the outer disk outflow is still 
dynamically evolving,
thus changing the cross-jet force equilibrium and forcing the inner 
structure to adjust accordingly.
The flare events are accompanied by a temporal change in outflow mass flux
and momentum.
Figure \ref{mflux} shows the mass loss rate in axial direction integrated
across the jet.
We see two flares with a 10\%-increase in the mass flux followed by
a sudden decrease of mass flux by a factor of two.
This behavior is also mirrored in the poloidal velocity profile.

Considering the ejection of large-scale flares and the follow-up
re-configuration of outflow dynamics,
we hypothesize that the origin of jet knots is triggered by such
flaring events.
Our time-scale for flare generation is of 1000 rotational 
periods and longer than the typical dynamical time at the jet base,
but similar to the observed knots.
The flare itself for about 30-40 inner disk rotation times.

\end{document}